# Femtosecond Thermal and Nonthermal Hot Electron Tunneling inside a Photoexcited Tunnel Junction

N. Martín Sabanés[1,2,†], F. Krecinic[1,†], T. Kumagai[1,3], F. Schulz[1], M. Wolf[1] and M. Müller[1,*]

[1]*Department of Physical Chemistry, Fritz Haber Institute of the Max Planck Society, Faradayweg 4-6, 14195 Berlin, Germany.*

[2]*IMDEA Nanoscience, Faraday 9, 28049 Madrid, Spain.*

[3]*Center for Mesoscopic Sciences, Institute of Molecular Science, 444-8585 Okazaki, Japan*

† Equal contributions

* Corresponding author: m.mueller@fhi-berlin.mpg.de

**Efficient operation of electronic nanodevices at ultrafast speeds requires understanding and control of the currents generated by femtosecond bursts of light. Ultrafast laser-induced currents in metallic nanojunctions can originate from photo-assisted hot electron tunneling or lightwave-induced tunneling. Both processes can drive localized photocurrents inside a scanning tunneling microscope (STM) on femto- to attosecond time scales, enabling ultrafast STM with atomic spatial resolution. Femtosecond laser excitation of a metallic nanojunction, however, also leads to the formation of a transient thermalized electron distribution, but the tunneling of thermalized hot electrons on time scales faster than electron-lattice equilibration is not well understood. Here, we investigate ultrafast electronic heating and transient thermionic tunneling inside a metallic photoexcited tunnel junction and its role in the generation of ultrafast photocurrents in STM. Phase-resolved sampling of broadband THz pulses via the THz-field-induced modulation of ultrafast photocurrents allows us to probe the electronic temperature evolution inside the STM tip, and to observe the competition between instantaneous and delayed tunneling due to nonthermal and thermal hot electron distributions in real time. Our results reveal the pronounced nonthermal character of photo-induced hot electron tunneling, and provide a detailed microscopic understanding of hot electron dynamics inside a laser-excited tunnel junction.**

**Introduction.** Hot carriers harvested in solids via optical excitation can drive a wide range of physical and chemical processes, the understanding of which lies at the heart of hot carrier science and technology.[1] The ultrafast dynamics of photoexcited hot carriers is intimately related to optoelectronic nanodevice operation. A particular example is laser-excited ultrafast scanning tunneling microscopy (STM), which pursues the ultimate goal to image the ultrafast dynamics of individual atoms and molecules on surfaces with simultaneous Angstrom spatial and femtosecond temporal resolution. Recent advances in combining ultrashort optical or terahertz (THz) pulses with low-temperature STM demonstrated the potential of light-driven STM to reach that goal[2–6]. Attaining ultrahigh spatiotemporal resolution in STM requires the generation of ultrafast, atomically localized tunneling currents. In the weak excitation regime, photo-assisted tunneling of hot electrons below the vacuum barrier gives rise to localized ultrafast currents that can be used for imaging in photon-driven STM[6–8]. In contrast, lightwave-driven STM operates in the strong-field regime, where an intense localized light field of several V/nm modulates the junction barrier and, hence, the tunneling current on sub-cycle time scales.[2,4,5,9–13] The two regimes are commonly distinguished by the Keldysh parameter $\gamma = \sqrt{\phi \, F_l/4\pi\omega}$, relating the work function of the material, $\phi$, to the photon energy $E_{\text{ph}} = \hbar\omega$ and the light field $F_l$. [14] Both processes are distinctly different in terms of the involved electronic excitation: whereas photo-assisted tunneling employs the tunneling of 'hot' but nonthermal electrons, lightwave-driven STM is mediated by adiabatic tunneling of thermal but 'cold' electrons directly from the Fermi level. Hereby, the terms 'hot' and 'cold' describe whether there is excess energy (e.g. via absorption) stored in the electronic system, whereas the terms 'thermal' and 'nonthermal' state whether the electronic distribution follows Fermi-Dirac statistics. In comparison to these well-established regimes, the ultrafast generation and tunneling of hot but thermalized electrons is rarely considered in the description of ultrafast photo-induced tunneling currents in STM, even though ultrafast electronic heating accompanies any pulsed laser-driven STM.

Focusing ultrashort laser pulses onto a metallic STM tip gives rise to light absorption and the generation of energetic, nonthermal electrons. These energetic carriers rapidly redistribute their energy via electron-electron and electron-phonon scattering, leading to the formation of a thermalized carrier distribution on the 10-100 fs time scale. Both, electrons from thermal and nonthermal distributions can tunnel through the potential barrier in STM, though with different time scales and energy distributions of the respective tunneling currents. Although the ultrafast dynamics and thermalization of hot electrons in metals has been intensively investigated,[15,16] the transient tunneling of thermalized hot electrons and their role in photoexcited STM is not well understood. However, ultrafast thermionic tunneling should coexist with both photon-driven and lightwave-driven tunneling currents. Moreover, the dynamics of hot electrons determines the initial step of laser heating of STM tips illuminated by ultrashort laser pulses[17,18]. It is thus of general interest to understand the role of hot electron dynamics for femtosecond heating and tunneling in photoexcited STM.

Transient thermionic emission has been discussed in the context of ultrafast electron guns using laser-excited nanotips.[19–22] It describes the emission of electrons above the vacuum barrier originating from the high-energy tail of a time-dependent Fermi-Dirac carrier distribution at elevated temperatures[23,24], and can be described by the Richardson-Dushmann equation.[25] In the presence of moderate electric fields $<10^8$ V/m, the barrier is lowered by the Schottky effect and the process is often termed Schottky emission or field-enhanced thermionic emission. Thermionic tunneling through the barrier, in comparison, can occur at much lower electron temperatures, but requires a large tunneling probability, i.e., due to the presence of very high fields >$10^8$ V/m or nanometer-sized tunneling gaps. In the case of biased nanotips this regime has been termed thermally-enhanced field emission,[19,26,27] and can be described by Murphy's and Good's equation for thermo-field emission.[28] In contrast to freestanding tips, STM junctions can facilitate the tunneling of hot electrons even without any external field. To understand thermionic tunneling on femtosecond time scales, the ultrafast evolution of the electronic temperature after laser excitation has to be known. Before electron-phonon equilibration has been achieved (>1 ps), different temperatures are assigned to the electronic and phononic subsystems. Their respective time evolutions can be described by the two-temperature model[15,19] (TTM) or by solving Boltzmann transport equations[16,20] for the coupled electron-phonon system. Whereas photo-assisted tunneling and optical-field-driven tunneling are temporally confined to the presence of the laser pulse, thermionic tunneling can occur on timescales longer than the laser pulse duration, and might lead to significantly delayed photocurrents in STM due to electron dynamics inside the tip.

Here we study the interplay between ultrafast thermal and nonthermal tunneling from a femtosecond laser-excited STM tip. By tracking the temporally delayed response of thermionic currents in real-time we can discriminate between non-instantaneous thermal and instantaneous nonthermal currents, without the need to distinguish their energy distributions as for example in photoemission experiments.[20,21,29] Energy information could, in principle, be obtained in STM by varying the bias voltage, but the bidirectional tunneling of photocarriers between tip and sample complicates interpretation of the respective hot carrier distributions. Instead, we measure the temporal response of ultrafast thermionic currents from a tungsten STM tip via phase-resolved sampling of single-cycle ultrabroadband THz pulses inside the tip-sample junction.[30] Specifically, we measure the time-dependent THz-induced change of photocurrents from the laser-excited STM tip[30–32]. The time dependence of the photocurrent change is determined by the instantaneous THz bias $U_{\text{THz}}(t)$ and the time evolution of the photocurrent $j_{\text{ph}}(t)$. In the case of prompt currents, the THz-field-induced photocurrent modulation yields the tip-enhanced THz near-field waveform. In contrast, delayed photocurrents that are not instantaneous on the time scale of the THz field will suppress higher THz frequencies and modify the measured waveform similar to a low-pass filter that limits sampling bandwidth. Knowledge of the original instantaneous THz waveform combined with numerical simulations of the time-dependent tunneling process allow us to extract the thermionic photocurrent dynamics and relate it to the electronic temperature evolution inside the STM tip. Whereas thermionic currents dominate the THz response of nearly freestanding tips, we find that nonthermal tunneling exceeds the thermal contribution for nanometer gap sizes. Even though nonthermal currents are prompt on the time scale of the THz field, and hence their temporal profile cannot be directly measured here, the distinct scaling of nonthermal and thermal tunneling with nanometer gap size allows us to unravel their relative contributions. Our results demonstrate that the increased tunneling rate in a nanoscale STM junction favors nonthermal tunneling over thermionic tunneling, despite the considerable increase of the electron temperature inside the tip caused by the increasing laser field enhancement at reduced gap size.

**Phase-resolved THz sampling of ultrafast photocurrents in STM.** Figure 1a) shows the experimental scheme of our measurement. We excite a grounded tungsten STM tip with 800 nm near-infrared (NIR) laser pulses of 10 fs duration with peak intensities ranging from $0.8 \cdot 10^{11}$ W/cm$^2$ to $2.5 \cdot 10^{11}$ W/cm$^2$. A positive bias of $U_{dc} = 8\,V$ is applied to the sample to suppress photocurrents from the sample with opposite sign and to ensure unidirectional flow of photocurrents originating from the tip. Phase-stable single-cycle broadband THz pulses are focused into the STM junction at variable delay $\tau$ with respect to the NIR pulses. The quasi-instantaneous field of the tip-enhanced THz field acts as a transient bias modulating the potential barrier and hence the current on femtosecond time scales. Provided that the THz-induced change of the photocurrent, $\Delta j_{THz}$, is dominated by a process that is instantaneous on the time scale of the THz field, the THz bias can be sampled with a time resolution as determined by the NIR pulse duration. The delay-dependent $\Delta j_{THz}(\tau)$ reveals the transient THz bias, and the local slope of the photocurrent-voltage curve allows to calibrate the THz bias amplitude $U_{THz}(t)$.[30,32] Precise knowledge of the waveform and amplitude of the THz bias will be essential to extract the time evolution of ultrafast non-instantaneous currents photoexcited inside the metallic junction, which will lead to temporal distortions of the measured THz waveform as illustrated in Figure 1a). We point out the difference to 'state-selective' THz-gated tunneling,[2] where the resonant THz-induced tunneling into individual molecular states during the peak of a THz half-cycle yields a time resolution faster than the THz half-cycle. Due to the threshold character of the resonant tunneling process, this regime does not require detailed knowledge of the THz waveform to access sub-cycle dynamics. In contrast, accessing the dynamics of 'non-resonant' systems, i.e., systems with a continuous variation of electronic states such as metals or semi-metals, on THz-sub-cycle time scales demands phase-resolved sampling of the THz waveform and knowledge of the THz bias transient.

The different photocurrent mechanisms that can be operative for a photoexcited STM tip are displayed in Figure 1b). At nanometer gap sizes, photo-assisted tunneling[6,7,33] through the barrier from a nonthermal step-like electron distribution can occur at energies that are multiples of the NIR photon energy $E_n = n\hbar\omega$. In comparison, the tunneling of thermalized hot electrons will be energetically distributed according to the Fermi-Dirac distribution at a given electron temperature $T_{el}$. At large gap size, electron emission above the barrier from either nonthermal multiphoton photoemission[17,22,29] or transient thermionic emission[19] will dominate the photocurrent. At high laser intensities, photocurrents might be generated in the strong-field regime, for which above-threshold photoemission[34–36] and optical-field-induced tunneling[4,37–40] have to be considered as well. The time-dependent THz field acts an additional quasi-static bias that modulates the NIR-induced photocurrent channels according to their respective dependence on the potential barrier between tip and sample. The THz bias, thus, selectively probes those photocurrent channels that are most sensitive to the potential barrier.

**Laser-induced ultrafast electron heating of the STM tip.** Figure 2a) shows the THz bias applied to the junction of a tungsten tip and Ag(111) surface, measured for weak optical excitation far from the tunneling regime at 1 μm tip-sample distance, where the photocurrent is dominated by above-barrier multiphoton photoemission. The peak THz voltage of $\sim 0.47\ V$ is small compared to the DC bias of 8 V to ensure that the THz field only slightly perturbs the potential barrier and to prevent quasi-static THz-induced tunneling. The photocurrent-voltage curve used for calibration of the THz bias is shown in Figure S.1 in the supporting information. Increasing the laser intensity leads to a broadening and continuous shift of the measured THz waveform $\Delta j_{\text{THz}}(\tau)$, as shown in Figure 2b), where the effect is most pronounced in the first half THz cycle. The observed waveform changes are very similar to those expected from a low-pass filter, evident also from the Fourier spectra shown in Figure 2c). Figure 2d) shows the laser intensity dependence of the time-averaged photocurrent measured with no THz bias applied, where the vertical purple-shaded area indicates the threshold intensity above which waveform deformations are observed for this particular tip. At low intensity, the photocurrent follows a nonlinear power scaling with effective nonlinearity $n = 3.4$, indicating that the photocurrent predominantly originates from multiphoton photoemission.[29] At laser intensities $I_{\text{peak}} \gtrsim 1.4 \cdot 10^{11}$ W/cm², a significantly reduced nonlinearity of $n \sim 0.9$ is observed, indicating transition to the strong-field regime.[37,41,42] The intensity at which this transition occurs is independent of the number of electrons per pulse (see Figure S.2 in the supporting information), implying that space charge effects are not responsible for the decreasing slope. Similar data from another tip are shown in Figure S.3 in the supporting information, where pronounced waveform changes are observed purely in the multiphoton photoemission regime. The observed waveform changes thus emerge in both the weak- and strong-field regime of photoemission. In both regimes, the photoemission process is prompt and the emitted current will be temporally confined to the laser pulse width. Hence, photoemission cannot explain the observed low-pass filter behavior, which requires a 'slow' photocurrent contribution, i.e., a delayed carrier response. We conclude that an additional delayed photocurrent component, which coexists with instantaneous weak- and strong-field photocurrents, is responsible for the observed waveform changes. As will be discussed below, the observed waveform changes can be assigned to delayed ultrafast thermionic currents from the STM tip, whose electronic subsystem is heated on the femtosecond time scale.

**Theoretical model.** To understand the observed waveform changes we calculate the photocurrent from the tip and its modulation by the THz field. The microscopic picture of our model and the electron dynamics inside the metal tip is depicted in Figure 3a). The electronic system is composed of two subsystems: (i) nonthermal electrons that follow a step-like distribution, and (ii) thermal electrons that follow a Fermi-Dirac distribution, as illustrated by the blue and red distributions in Figure 1b), respectively. Photoexcitation by an ultrashort laser pulse initially promotes electronic single-particle excitations, leading to a nonthermal electron distribution that thermalizes via electron-electron scattering with a rate $\propto 1/\tau_{e-e}$, which takes place on time scales of typically few 10 fs in transition metals.[15,16,43] On longer time scales of ~1 ps, electron-phonon scattering leads to energy transfer from the electronic to the phononic subsystem with a rate $\propto 1/\tau_{e-ph}$.[15,16] Due to the different time scales of $\tau_{e-e}$ and $\tau_{e-ph}$, it is reasonable to assume that electron-phonon coupling only leads to cooling of the thermal electron distribution, but does not remove energy from the nonthermal part of the electronic system. In addition, ballistic transport and diffusion of hot electrons into the bulk can influence the electron distributions at the tip surface.[15,44] Both, nonthermal and thermal electron distribution, can give rise to a photocurrent from the tip, which we write as the sum

$$j_{ph}(t) = j_{th}(t) + \delta(t - t_0) j_{nonth} \qquad (1)$$

of an ultrafast time-dependent thermionic current $j_{th}(t)$ and an instantaneous nonthermal current $j_{nonth}$, where $\delta(t - t_0)$ is a delta function located at the laser pulse center $t_0$. The separation in equation (1) and the assumption of an instantaneous nonthermal current is reasonable here considering the limited time resolution of 10's fs determined by the finite THz bandwidth, which is insufficient to resolve the electron thermalization process via electron-electron scattering.

The time-dependent thermionic current through the THz-modulated potential barrier is calculated from the total charge $Q_e$ extracted per pulse due to a thermalized hot electron distribution,

$$j_{th}(\tau) \propto Q_e(\tau) = e \int D[W, F(t - \tau)] \, N_{th}[E, T_{el}(t)] \, dW dE dt, \qquad (2)$$

where $\tau$ is the delay between THz and NIR pulse, $F(t) = F_{dc} + F_{THz}(t)$ is the combined DC and THz electric field, and $D$ is the time-dependent transmission probability, where $W = E - (p_x^2 + p_y^2)/2m_e$ is the kinetic energy of the electron in the positive z-direction.[45] $N_{th}$ is the time-dependent thermal electron occupation at energy $E$ and temperature $T_{el}$ which follows the Fermi-Dirac distribution

$$N_{th}(E, T_{el}(t)) = \frac{1}{\exp[(E - E_F)/k_B T_{el}(t)] + 1}. \qquad (3)$$

The THz-modulated transmission probability $D$ is calculated by solving the one-dimensional Schrödinger equation

$$-\frac{\hbar}{2m_e}\frac{d^2\psi}{dz^2} + V(F, z)\psi = W\psi \qquad (4)$$

for the time-dependent quasi-static potential

$$V(t,z) = \begin{cases} 0 & z \leq 0 \quad (5a) \\ E_F + \phi - \frac{e^2}{4z} - eF(t)z & z > 0 \quad (5b) \end{cases}$$

where $\hbar$ and $m_e$ are the reduced Planck constant and electron mass, respectively, $E_F$ is the Fermi level, $\phi$ is the work function of the tip and the third term in $(5b)$ describes the effect of image charges. Due to the high positive sample bias we can neglect current contributions originating from the sample in the present calculations. The time evolution of the electronic temperature $T_{el}(t)$ is obtained by solving the two-temperature model (TTM) in three dimensions inside the tip including ballistic and diffusive electron transport using COMSOL Multiphysics (details are described in section 4 of the supporting information). The TTM source term $S_b$ for electronic heating is obtained from the spatially inhomogeneous absorbed power density, convoluted with the mean free path of electrons in tungsten[46] to account for the initial fast redistribution of energy due to ballistic transport. The left panel in Figure 3b) shows the spatial profile of $S_b$ inside the tip after ballistic energy redistribution, which on the nanometric scale of the tip occurs instantaneously on the time scale of the THz field. For a given tip-sample geometry, we obtain the ultrafast electronic temperature evolution, the locally enhanced optical field as well as the DC field along the surface of the tip using COMSOL simulations. The middle and right panel in Figure 3b) show the electronic temperature distribution inside the tip at $\tau = 20$ fs and $\tau = 1$ ps, respectively. $T_{el}$ initially resembles the profile of $S_b$ but becomes blurred on longer time scales due to diffusive hot carrier transport. Finally, the experimental calibration of the THz bias allows us to fix the ratio of DC and THz field. As we know the waveform of the THz bias, we can thus calculate the THz-induced change of the ultrafast thermionic current $\Delta j_{th}(\tau) = j_{th}(\tau) - j_{th}(\tau < 0)$ by calculating the emitted charge in the presence of the instantaneous THz field using equation (2).

The quasi-instantaneous photocurrent due to nonthermal electrons can be written as a superposition of currents through each multiphoton channel $n$,

$$j_{\text{nonth}} = C_0 \sum_n j_n \quad \text{with} \quad j_n = D_n N_n = D_n \left[(C_{\text{ex}})^n |F_{\text{IR}}|^{2n}\right], \tag{6}$$

where $F_{\text{IR}}$ is the tip-enhanced laser field, $D_n$ is the transmission probability at energy $E_n = E_F + n\hbar\omega$ with $n = [1,2,3,...]$, and $C_0$ is a constant scaling factor. The excitation constant $C_{\text{ex}} < 1$ is a free parameter whose dependence $(C_{\text{ex}})^n$ accounts for the decreasing probability for excitation into higher multiphoton channels. However, the relative contribution from higher channels increases with increasing laser intensity as expected due to $|F_{\text{IR}}|^{2n}$. The THz-induced change of the nonthermal current, $\Delta j_{\text{nonth}}(\tau)$, is calculated from equation (6) taking into account the time dependence of the transmission probability $D_n$ due to the instantaneous THz field at the tip surface. Finally, the simulated waveforms are obtained from the sum of the THz-induced change of the photocurrent $\Delta j_{\text{THz}}(\tau) = \Delta j_{\text{th}}(\tau) + \Delta j_{\text{nonth}}(\tau)$. The free parameters of our model are the incident laser intensity, the constants $C_0$ and $C_{\text{ex}}$, and the tip-sample geometry including the gap distance $d$.

At high laser fields approaching $\gamma_{\text{IR}} \sim 1$, above-threshold photoemission becomes significant. Yet, as those channels are relatively insensitive to the potential barrier and, hence, the THz bias, we neglect them here and restrict our calculations to channels below or close to the top of the barrier ($n \leq 3$ in our case). At even higher laser fields, entering the optical tunneling regime at $\gamma_{\text{IR}} \ll 1$, the laser field is significantly stronger than the applied THz and DC fields. Hence, optical tunneling should be almost unaffected by the weak THz fields applied here, and, therefore, the THz-induced change in optical-field-driven tunneling currents is expected to be very small. We simulated NIR-lightwave-tunneling from the THz-biased tip by adding the NIR laser field to the total field $F(t)$ in equation (5b). The results confirmed that optical tunneling contributes insignificantly to the THz-induced change of the photocurrent due to its low bias sensitivity in the parameter range used here. We can thus neglect strong-field effects and optical tunneling in our waveform simulations. We further note that the measured photocurrent is NIR-cycle-averaged and integrated over energy. In contrast to energy-resolved measurements, which record the final kinetic energy of the photoelectrons after propagating through the oscillating THz field,[31] our measurement is not sensitive to electron propagation but only to the instantaneous THz field at the tip surface. Since $F_{dc} \gg F_{THz}$ we can also neglect THz streaking of photoelectrons back into the tip.

**Simulation of ultrafast thermionic currents and THz waveforms from a photoexcited STM tip**. Figure 3c) shows calculated THz waveforms obtained at the apex of a tip with work function $\phi = 4.5$ eV, tip radius $R_{tip} = 5$ nm and shaft opening angle $\alpha_{tip} = 4°$ (a SEM image of the STM tip used in this work is shown in Figure S.5 in the supporting information). The simulations with incident laser intensities close to those estimated from our experimental conditions reproduce the measured waveform broadening and temporal shift very well. The exact laser intensities required to observe electronic heating in the simulated waveforms vary with tip size and shape, which is consistent with the experimental variation of the laser power we require to observe waveform deformations for different tip conditions. The corresponding time dependence of the electron temperature and thermionic current is shown in Figures 3d) and 3e), respectively. Electronic temperatures of several 1000

Kelvin are reached inside the tip on ultrafast time scales due to the small heat capacity of electrons in metals.[47] It is important to note that thermionic currents alone do not describe the waveforms sufficiently well, especially at low laser intensities at which $\Delta j_\text{th}(\tau)$ does not yield the original THz waveform (pure 'thermionic waveforms' are shown in Figure S.4.2 in the supporting information). To reproduce the laser power dependence of the waveforms in Figure 2b) precisely, the combination of thermal and nonthermal currents is required. Whereas nonthermal currents, at far distances predominantly emitted through $n = 3$, dominate and reproduce the original THz waveform at low laser intensities, delayed thermionic currents are responsible for the waveform distortions at higher laser intensities. Figure 3f) shows the calculated power scaling of the THz-induced thermionic current $\Delta j_\text{th}$ compared to the power scaling of the nonthermal current $\Delta j_{n=3}$ through channel $n = 3$. The slope of $\Delta j_\text{th}$ is much steeper than that of $\Delta j_{n=3}$ and continuously decreases at higher laser intensities, as opposed to the constant slope $n = 3$ of the nonthermal photocurrent. Whereas the scaling of $\Delta j_{n=3}$ is determined solely by the excitation into states at $E_{n=3} = 3\hbar\omega$, which increases as $|F_\text{IR}^2|^3$, the power scaling of $\Delta j_\text{th}$ is determined by the increasing electron temperature and its decreasing sensitivity to the barrier, and hence THz field, at high $T_\text{el}$. Overall, the relative contribution of $\Delta j_\text{th}$ is insignificant at low intensities, but increases rapidly with increasing laser intensity. It has been discussed previously that non-integer nonlinearities, which are frequently measured for photoemission from nanotips, can be caused by long-lived hot electron distributions inside the tip,[48,49] which is supported by our results. We note that we can exclude that the delayed currents originate from laser-driven rescattering[50] and delayed re-emission of electrons from the tip,[21] as the waveform distortions are observed also (and for some tips solely) in the weak-field regime, in which laser-driven electron scattering is negligible (section 3 of the supporting information). The results in Figures 2 and 3 thus confirm that femtosecond laser heating of the STM tip can lead to ultrafast thermal currents persisting on time scales of several 100 fs, i.e., much longer than the exciting laser pulse duration.

**Gap size dependence.** To examine the role of thermalized hot electrons for photo-induced tunneling in STM, we measure the THz waveform versus tip-sample distance as plotted in Figure 4a). The NIR power is set to a value that yields significant waveform deformations due to delayed thermal photocurrents at large tip-sample distances. The zero position $\Delta z = 0$ nm is defined by the STM set point of $U_{dc} = 10$ V and $j = 1$ nA, at which the current is dominated by static tunneling. The relative gap distance is varied in the range between $-0.2$ nm and 6 nm (negative values mean smaller gap size). We can roughly estimate the absolute tip-sample distance by fitting the decay of the static current (dashed line in Figure 4b)) and extrapolating to the quantum conductance $G_0$, which yields $d \sim 2.9$ nm for $\Delta z = 0$ nm, which is reasonable at our conditions. As demonstrated previously, the THz bias does not depend on gap size,[11,30] and therefore the THz bias measured at large gap distance (grey dashed line) applies to all distances. We note that 'static' rectified THz currents are negligible due to the small THz bias amplitude and the comparably large gap distance when operating at 10 V bias. Starting from a distorted waveform at $\Delta z = 6$ nm, the waveform continuously transforms such that it reproduces the original waveform of the THz bias at the smallest gap size. Complete vanishing of the thermal waveform distortions correlates with the onset of static tunneling, as evident from the distance dependence of the current plotted in Figure 4b). This is surprising, as naively one would expect a slower current decay due to increased tunneling at lower energies, for which the occupation and current decay times are larger than at the high-energy tail of the Fermi distribution. In addition, ultrafast heating will be enhanced at nanometer distances, because the optical field, and hence the absorbed power, are inversely proportional to the gap size. The dependence of the waveform deformations on the gap distance, which is distinctly different from the dependence on laser power shown in Figure 2b), thus clearly indicates that instantaneous current contributions must exceed thermionic tunneling at reduced gap size close to the static tunneling regime. Figure 4c) shows the effective nonlinearity extracted from current-distance curves measured at varying laser pulse energy (details are described in section 6 in the supporting information). Whereas at large tip-sample distances the photocurrent exhibits a nonlinearity of $n_{\text{eff}} \sim 3$, the nonlinearity continuously decreases with decreasing $\Delta z$, indicating that photo-assisted tunneling of lower orders dominates at small gap sizes.

The observations in Figure 4 suggest that at close distances the instantaneous tunneling of nonthermal electrons competes with delayed thermionic tunneling from the laser-excited STM tip. To corroborate this assumption, we simulate THz waveforms for varying gap distances $d$. With decreasing $d$, the DC field, THz field, and optical field at the tip apex increase, as well as the absorbed power and hence the electronic temperature inside the tip. Figures 5a) and 5b) show the distance scaling of $T_{\text{el}}$, $F_{\text{dc}}$ and the optical field enhancement at the tip apex, together with the decay time $\tau_{th}$ of the ultrafast thermionic current. The longer decay times $\tau_{th}$ at reduced gap distances result from the higher electron temperatures and an increasing contribution from electrons tunneling at lower energies. Figure 5c) compares the gap size dependence of the THz-induced currents $\Delta j_{\text{THz}}$ due to thermal and nonthermal electrons of orders $n = [1,2,3]$, respectively. The THz waveforms calculated from the superposition of those currents are plotted as a function of $d$ in Figure 5d), which reproduces the experimental observations in Figure 4a) well. We can thus explain the gap size dependence of the measured THz waveforms by the competition between delayed thermal and prompt nonthermal tunneling from the photoexcited STM tip. The distinctive THz waveform changes can be understood from the distance-dependent mixture of thermal and nonthermal photocurrents, whose relative contributions vary with gap size according to their specific tunneling probabilities and nonlinear dependence on the local laser field, as reflected in the different slopes in Figure 5c). Whereas the nonthermal contribution from $n = 3$ increases mainly due to its nonlinear dependence on local laser intensity, the lower order contributions $n = 2$ and, in particular, $n = 1$ increase rapidly at reduced gap size due to their steeply increasing tunneling rates. In contrast, the source term of electronic heating in the TTM scales linearly with laser intensity. Moreover, the thermionic current is integrated over energy and thus exhibits an overall reduced sensitivity to the barrier compared to low-order nonthermal tunneling channels. This becomes apparent from the energy distribution of $j_{\text{th}}$ plotted in Figure 5e) together with the corresponding thermal electron occupation $N_{\text{th}}$ and transmission probability $D$ for three gap sizes at $\tau = 20$ fs. Even though the peak of the thermionic current distribution shifts to lower energies and develops a significant low-energy tail at nanometer gap distances, the majority of thermalized electrons is emitted at energies ~2 eV above $E_F$ and, hence, experiences a reduced sensitivity to the THz field compared to the nonthermal electrons excited into channel $n = 1$. It is the nonlinear laser intensity dependence of channels $n > 1$ on the one hand, and the higher sensitivity of low-order channels to the tunneling barrier on the other hand, which causes the instantaneous nonthermal currents $\Delta t_{\mathrm{nonth}}$ to exceed the delayed thermionic current $\Delta j_{\mathrm{th}}$ at nanometer gap sizes. Our findings clearly show the pronounced nonthermal and prompt character of hot electron-induced tunneling from a photoexcited STM tip.

Finally, we note that our simple model can reproduce the data reasonably well without calculation of the full nonthermal step-like distribution. The experimental data, on the other hand, shows a more rapid waveform change, within a distance variation of only 1 Å close to the set point, than that observed in the simulations. This indicates that our model might underestimate nonthermal hot electron tunneling close to the Fermi level, which exhibits an even higher sensitivity to the tip-sample distance. Further improvement of the agreement between the experimental and simulated results could be obtained by extending our model by (i) the simulation of the electron thermalization process, (ii) the inclusion of the full nonthermal distribution, and (iii) replacing the Fowler-Nordheim barrier by the Simmons barrier,[51] which is more appropriate at nanometer gap sizes. We expect that more detailed spectroscopic information about the nonthermal distributions might be obtained in future experiments by variation of the DC and THz bias. Moreover, it may be possible to measure the electron thermalization process and its effect on tunneling by using shorter THz transients to increase the time resolution.

In conclusion, we investigated the competition between ultrafast thermal and nonthermal photocurrents inside a laser-excited STM junction via phase-resolved sampling of broadband THz transients. Our results reveal that hot electron tunneling from a laser-excited STM tip is dominated by nonthermal electron distributions, accompanied by delayed tunneling of thermalized hot electrons. Outside the tunneling regime, such thermalized hot electrons dominate the THz-induced photocurrent from a tungsten tip excited at high laser intensities. However, this current contribution is exceeded by nonthermal currents when entering the tunneling regime. We infer the markedly nonthermal character of the photoexcited electron distribution from the distinctly different durations of the thermal and nonthermal photocurrents. Prompt nonthermal tunneling is followed by a thermionic photocurrent that can be delayed by several 100 fs, as determined by the time at which the laser-excited STM tip reaches its peak electronic temperature, which can increase up to several thousand Kelvin. The weak THz field serves as a highly sensitive and selective probe for photo-induced tunneling currents in ultrafast laser-excited STM, which is a unique capability of THz-gated STM. Detailed information about the phase and amplitude of the THz bias allows one to extract the ultrafast dynamics of hot electrons on time scales much shorter than a single THz-cycle even in non-resonant systems like metals or semi-metals. We envision that extension of our approach by the imaging capabilities of STM will provide a new route for the spatiotemporal investigation of ultrafast carrier dynamics at photoexcited surfaces on Angstrom length and femtosecond THz-sub-cycle time scales.

**Methods**

Measurements are performed with a customized STM from Unisoku (USM-1400 with Nanonis SPM controller) operated at room temperature and under ultrahigh vacuum conditions (base pressure of $< 5 \times 10^{-10}$ mbar). The spring-loaded STM platform is equipped with two off-axis parabolic mirrors (1× bare Au and 1× protected Ag, 1" diameter, 35 mm focal length) for illumination with VIS-NIR laser and broadband THz pulses, respectively. Both beams are incident to the tip axis at an angle of 68°. The THz beam enters the UHV chamber through a 500 μm thick diamond window and is focused by the Au PM. The Ag PM mirror is used to focus the VIS-NIR pulses for photoexcitation of the STM junction. Precise focus alignment on the tip apex is ensured by precise positioning of the PMs in UHV, which are motorized and can be moved in xyz-direction (Attocube GmbH). The tip position is fixed and the sample is moved for coarse motion and scanning. The DC bias is applied to the sample and the current is collected from the grounded tip. The current preamplifier (Femto DLPCA) is operated at a gain of $10^9$ V/A at 1 kHz bandwidth. A mechanical chopper operated at a frequency of 607 Hz modulates the power of the NIR laser beam used for THz generation, and lock-in detection is used to detect the THz-induced current. Repeated cycles of Ar+ sputtering and annealing up to 670 K were performed to clean the Ag(111) sample before measuring. Electrochemically etched tungsten tips are transferred to UHV immediately after etching.

The laser system is a broadband optical parametric chirped-pulse amplifier (Venteon OPCPA, Laser Quantum) delivering VIS-NIR laser pulses of 10 fs duration at 800 nm center wavelength and with 3 μJ energy (2 μJ are available for the THz-STM setup) at 1 MHz repetition rate. Part of the laser power is used for the generation of single-cycle THz pulses from a spintronic THz emitter (STE, 5.8 nm thick W/CoFeB/Pt trilayer on 500 μm sapphire substrate[52]) excited at normal incidence in transmission geometry. A motorized translation stage is used to control the delay between the THz pulses and the VIS-NIR laser pulses used for photoexcitation of the STM. Part of the VIS-NIR laser pulses can be overlapped collinearly with the THz beam for THz pulse characterization using electro-optic sampling, as well as for precise THz beam alignment inside the STM.

Numerical simulations are performed using COMSOL Multiphysics 5.6., which provide the required values of the DC field, tip-enhanced optical field, and electronic temperature evolution inside the tip. The photo-induced thermal and nonthermal currents and their THz-induced change (THz waveforms) are simulated using custom Python scripts. Details about the model are described in the supporting information.

………..……………

**AUTHOR INFORMATION**

**Corresponding Author**

*E-mail: m.mueller@fhi-berlin.mpg.de

**Notes**

The authors declare no competing financial interest.

**Acknowledgement**

The authors thank A. Paarmann, E. Ernstorfer and T. Kampfrath for valuable discussions. The authors further thank Unisoku Inc. for support and discussions in the development of the STM instrumentation. We thank T. Kampfrath, T. Seifert, G. Jakob and M. Kläui for providing us with the spintronic THz emitter, and H. Kirsch for tip preparation support. N.M.S. acknowledges the MSCA program MSCA-IF-2019-892667.

**Author contributions**

M.M. conceived the experiment. N.M.S. and M.M. performed the measurements and analyzed the data. F.K., N.M.S. and M.M. performed the numerical simulations. F.K. developed the python codes. T.K., M.M. and M.W. designed the STM instrumentation. N.M.S., F.S. and M.M. constructed the experimental setup. M.M. wrote the manuscript. All authors contributed to the discussion and commented on the manuscript.

**Supporting Information.** (1) Calibration of the THz bias, (2) Bias dependence of photocurrent-power scaling at large gap size, (3) THz waveforms for a tip operating solely in the weak-field regime, (4) Theoretical model, (5) SEM image of STM tip, (6) Power-dependent current-distance curves.

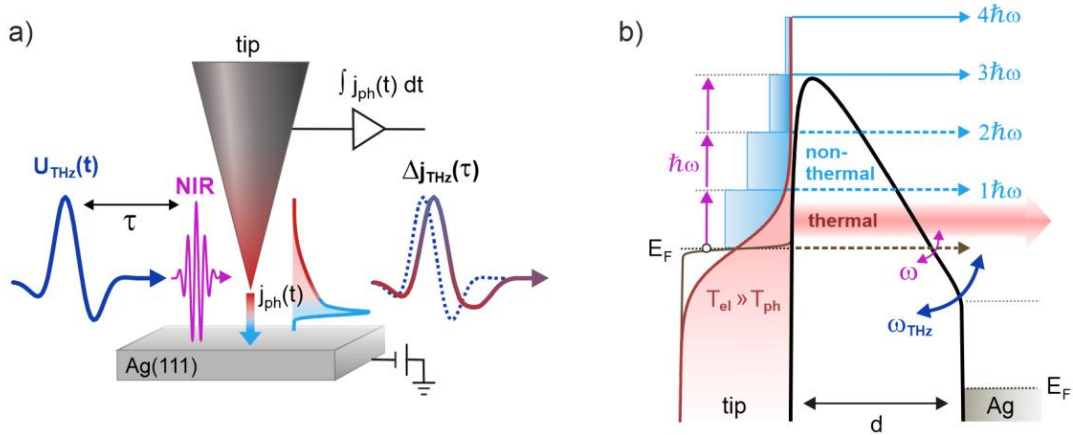

*Figure 1.* Real-time sampling of ultrafast thermionic currents in a photoexcited THz-gated STM junction. (a) A time-dependent photocurrent $j_{ph}(t)$, which can contain fast (light blue) and slow (red) components, is excited from the STM tip by 10 fs NIR laser pulses. The time evolution of the photocurrent is probed by phase-stable single-cycle THz pulses acting as a quasi-static bias that modulates the potential barrier. A delayed photocurrent response is encoded in the measured THz waveform obtained from the THz-induced change of the photocurrent, $\Delta j_{THz}(\tau)$. (b) Photocurrent channels from the laser-excited STM tip. On the time scale of the NIR pulse duration, above-barrier photoemission and photo-assisted tunneling at energies $E_n = n\hbar\omega$ (light blue arrows) lead to nonthermal currents. At longer times >10 fs, the tunneling of thermalized hot electrons (red arrow) can lead to delayed photocurrents depending on the time evolution of the electron temperature $T_{el}$. Strong NIR laser fields can induce sub-optical-cycle tunneling from a 'cold' electron distribution at the Fermi level (gray dashed arrow). The THz field modulates the potential barrier and the photocurrent yield of all channels according to their respective bias dependence and emission times.

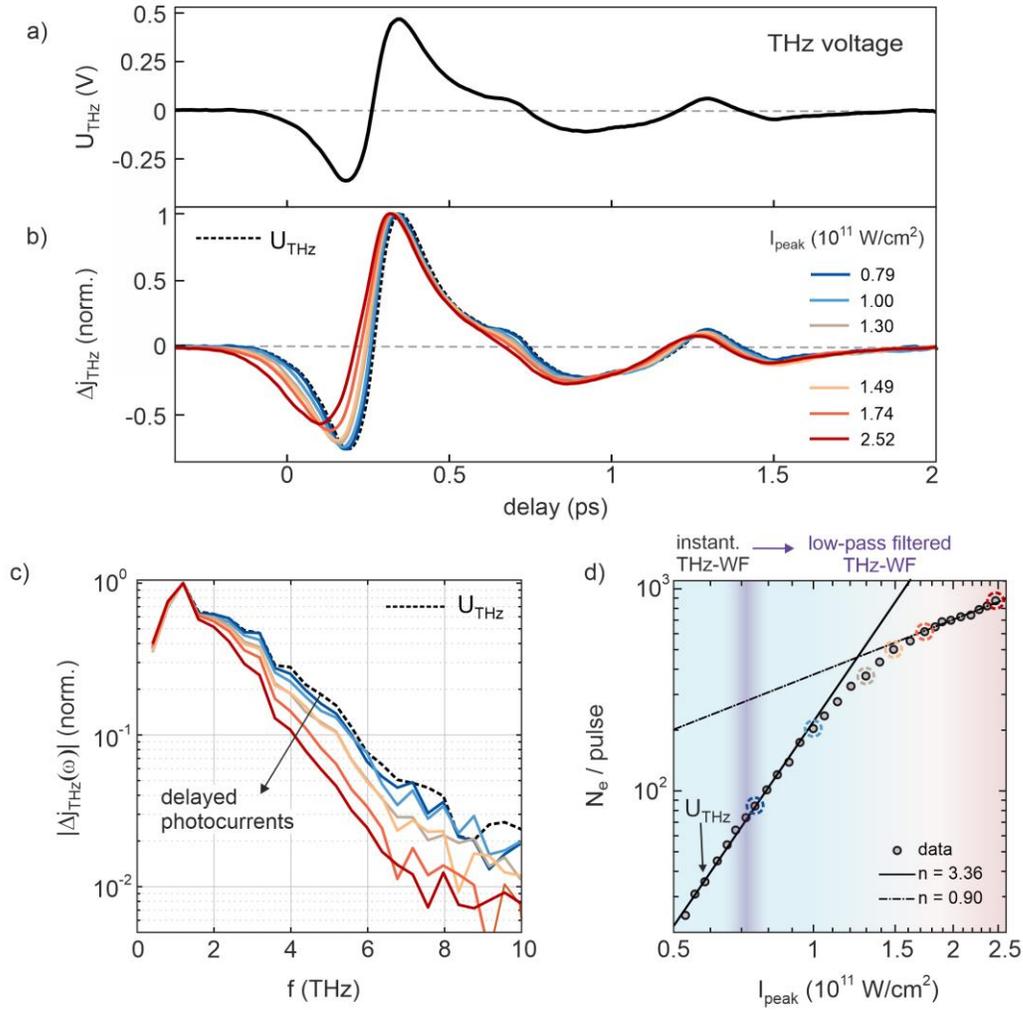

*Figure 2* Phase-resolved THz sampling of ultrafast non-instantaneous photocurrents from the STM tip at large gap distances. (a) THz bias measured at weak optical excitation ($I_{peak} = 0.65$ W/cm²), where quasi-instantaneous multiphoton photoemission above the barrier dominates the photocurrent. (b) THz waveforms measured at increasing NIR laser intensity, exhibiting a power-dependent distortion that resembles the characteristic of a low-pass filter, as also evident from the Fourier spectra of the waveforms plotted in (c). (d) The time-averaged photocurrent exhibits a transition from the multiphoton regime (power exponent $n = 3.36$) to the strong-field regime ($n = 0.9$). Low-pass filtered THz waveforms due to delayed photocurrents are observed in both regimes. ($d = 1$ μm, $U_{dc} = 8$ V)

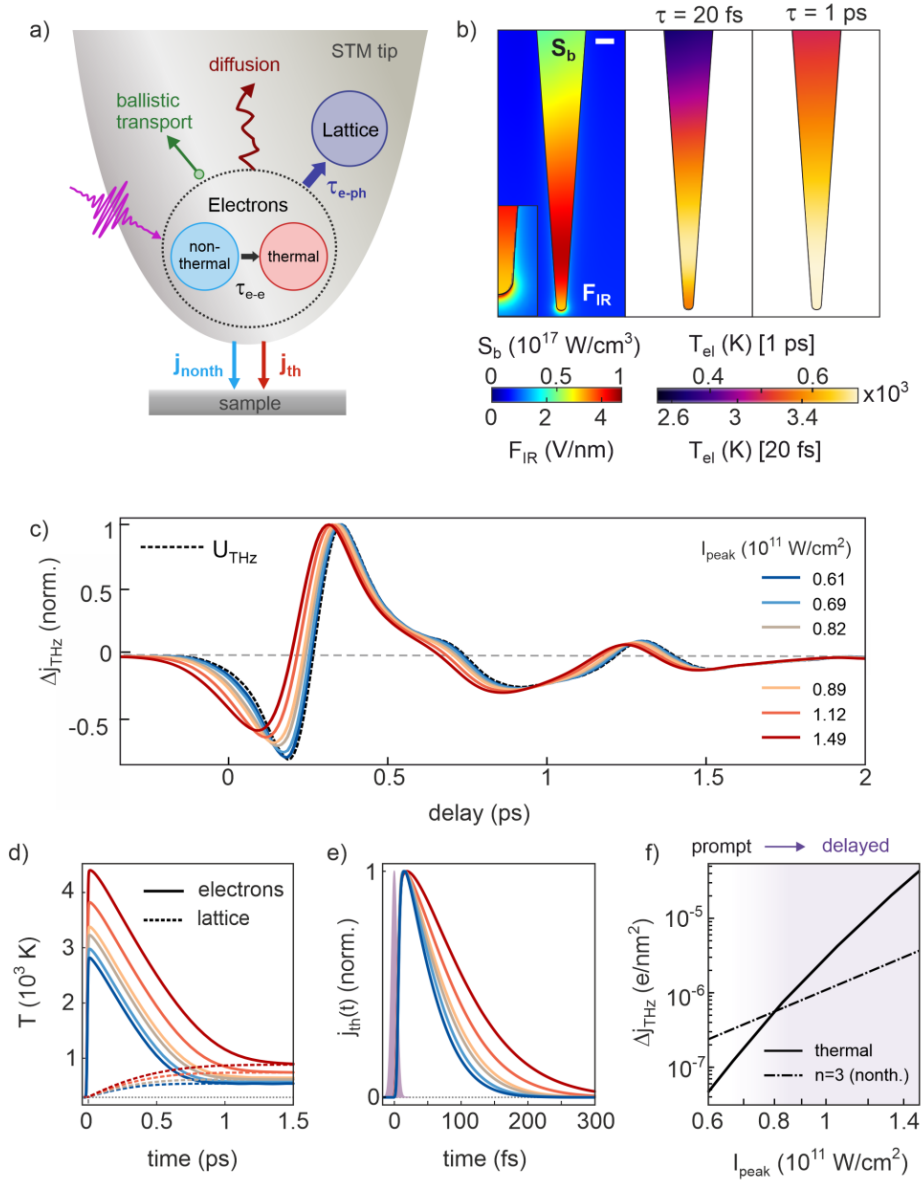

*Figure 3* Simulation of ultrafast thermionic currents and THz waveforms from a photoexcited STM tip. (a) Microscopic picture of electron dynamics inside the tip. (b) Left panel: cycle-averaged NIR peak electric field $F_{IR}$ (outside tip) and absorbed power density $S_b$ (inside tip) after ballistic distribution of absorbed energy. Middle and right panel: Electronic temperature distribution inside the tip after $\tau = 20$ fs and $\tau = 1$ ps. Scale bar is 20 nm. (c) Simulated THz waveforms retrieved from superposition of instantaneous nonthermal photocurrents and thermionic currents. (d) Time evolution of electron (solid) and lattice (dashed) temperatures and (e) the resulting thermionic current at the tip apex for different laser intensities. The shaded area in (e) shows the intensity envelope of the NIR pulse. (f) Calculated power scaling of the THz-induced thermionic current $\Delta j_{th}$ compared to that of the nonthermal current $\Delta j_{n=3}$ through channel $n = 3$. ($d = 1$ µm, $F_{dc} = 0.43$ V/nm at $U_{dc} = 8$ V, $F_{THz} = 0.025$ V/nm, $R_{tip} = 5$ nm, $C_{ex} = 10^{-22}$, $C_0 = 2 \cdot 10^{23}$). The free parameters $I_{peak}$, $C_{ex}$ and $C_0$ are adjusted to best fit the experimental THz waveforms in Figure 2 (b).

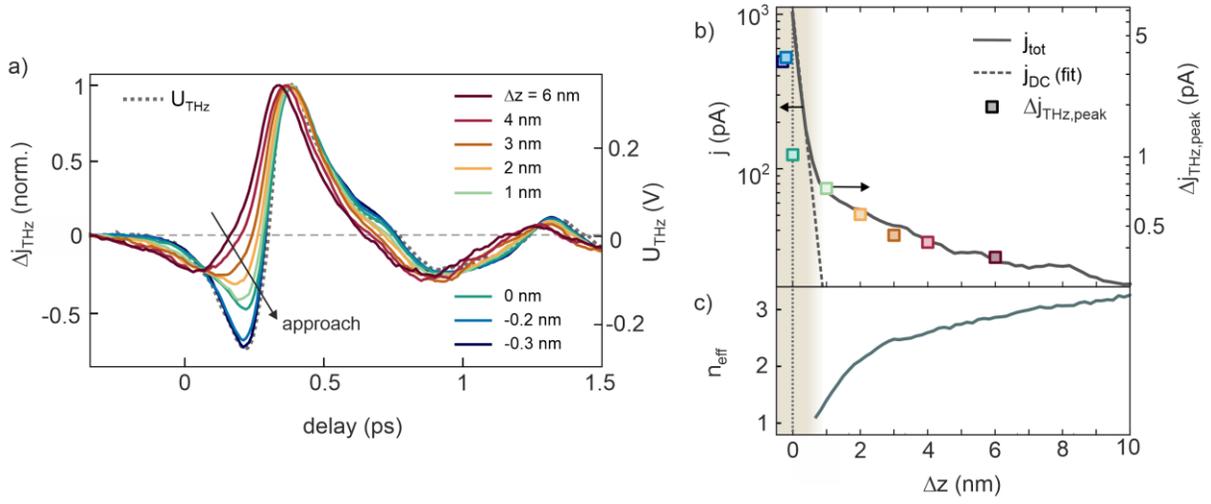

*Figure 4* Gap size dependence of measured THz waveforms. (a) THz near-field waveforms measured against relative gap distance $\Delta z$ from the set point. The THz waveform gradually transforms into the original THz bias waveform at short gap distances. (b) Gap size dependence of the total current (left y-axis) and THz-induced change of the photocurrent (right y-axis, peak value). Recovery of the original THz bias waveform correlates with the onset of static tunneling (dashed line). (c) Effective nonlinearity $n_{eff}$ of the photocurrent versus relative gap distance. (Set point 1 nA and 10 V, laser pulse energy $E_p = 4.3$ nJ).

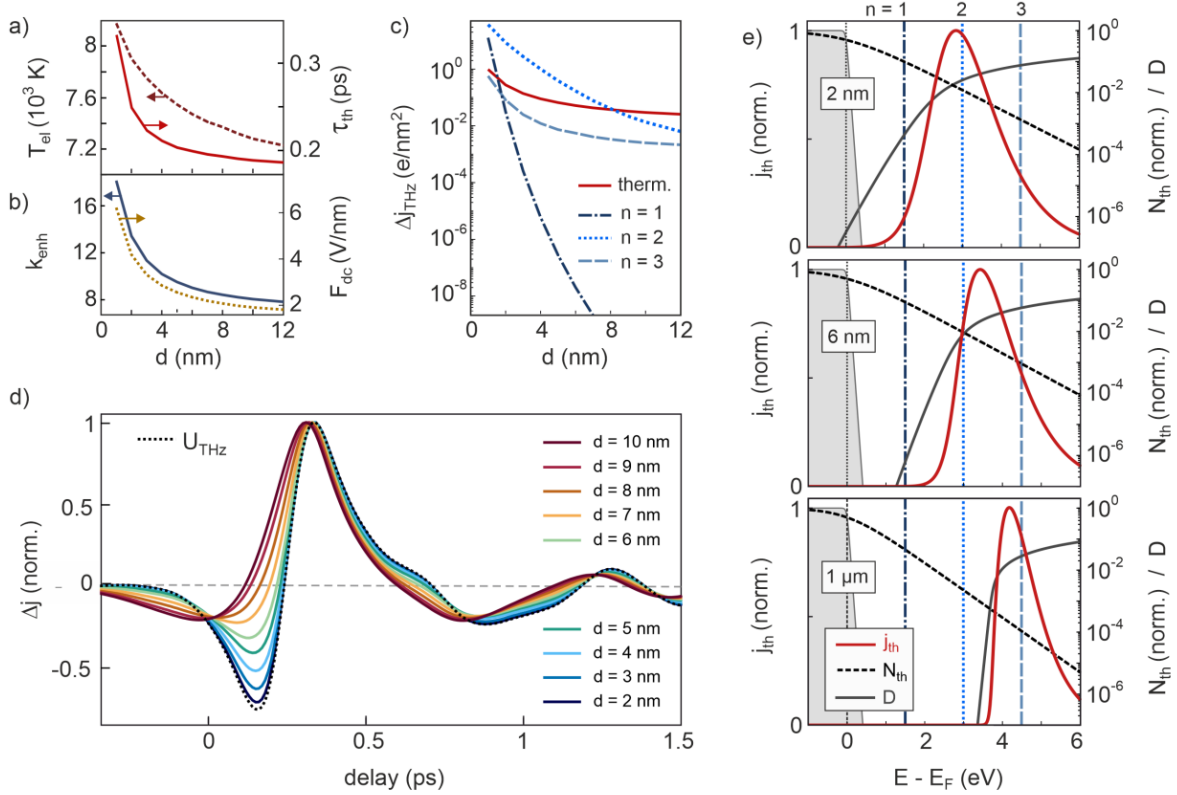

*Figure 5* *Simulation of photocurrent channels and THz waveforms versus tip-sample distance. (a) Distance scaling of electron peak temperature (left) and decay time of the thermionic current (right). (b) Distance scaling of optical field enhancement (left) and DC field (right). (c) Dependence of the THz-induced photocurrent change $\Delta j_{THz}$ on gap size for thermionic tunneling (red) and nonthermal photo-assisted tunneling at energies $E_n = n\hbar\omega$ (dashed). (d) THz waveforms calculated for the combined photocurrent channels shown in (c) at varying gap distance. (e) Energy distribution curves of the thermionic current (solid red, left y-axis) at delay $\tau = 20$ fs after photoexcitation for three gap distances 1 µm (bottom), 6 nm (middle) and 2 nm (top). The black dashed curves show the corresponding Fermi-Dirac-distributions ($N_{th}$, log-scale) with peak temperatures of $T_{el} = 5722$ K at 1 µm, $T_{el} = 7466$ K at 6 nm, and $T_{el} = 7901$ K at 2 nm. The black solid curve shows the transmission probability D in the absence of the THz field. The blue vertical lines mark the positions of the nonthermal photocurrent channels $n = [1,2,3]$. The gray shaded area shows the Fermi-distribution at room temperature.*

Supporting Information

# Femtosecond Thermal and Nonthermal Hot Electron Tunneling inside a Photoexcited Tunnel Junction

N. Martín Sabanés[1,2,†], F. Krecinic[1,†], T. Kumagai[1,3], F. Schulz[1], M. Wolf[1] and M. Müller[1,*]

[1]*Department of Physical Chemistry, Fritz Haber Institute of the Max Planck Society, Faradayweg 4-6, 14195 Berlin, Germany.*

[2]*IMDEA Nanoscience, Faraday 9, 28049 Madrid, Spain.*

[3]*Center for Mesoscopic Sciences, Institute of Molecular Science, 444-8585 Okazaki, Japan*

[†] Equal contributions

[*] Corresponding author: m.mueller@fhi-berlin.mpg.de

**Content**

1. **Calibration of the THz bias**

2. **Bias dependence of photocurrent-power scaling at large gap size**

3. **THz waveforms for a tip operating solely in the weak-field regime**

4. **Theoretical model**

5. **SEM image of STM tip**

6. **Power-dependent current-distance curves**

# 1. Calibration of the THz bias

The THz voltage between the tip-sample junction can be calibrated if (i) the NIR-induced photocurrent emitted from the tip is instantaneous on the time scale of the THz field, and if (ii) the current-voltage dependence of the photocurrent is known. Figure S.1a) shows the photocurrent versus DC bias measured for the same tip condition as the data shown in Figure 2) in the main manuscript. The tip-sample distance is 1 µm and the incident laser intensity $I_{\text{peak}} = 0.6x10^{11}$ W/cm² is low enough to ensure operation without thermionic current contributions. Due to kinetic excess energy of the emitted photoelectrons, we observe photocurrent from the tip also at moderate negative bias. At larger negative bias the photocurrent reverses sign and is dominated by photoemission from the Ag sample. Vice versa, photoelectrons from the sample might contribute to the current at moderate positive sample bias. Consequently, we record the THz waveform at 8 V DC bias. Due to the quasi-static nature of the THz field, the THz amplitude can be calibrated by dividing the THz-induced change of the photocurrent by the linear I-V slope obtained from the DC bias dependence. Figure S.1b) shows the retrieved THz bias waveform (see also Figure 2a)). The THz bias is calibrated routinely before each measurement.

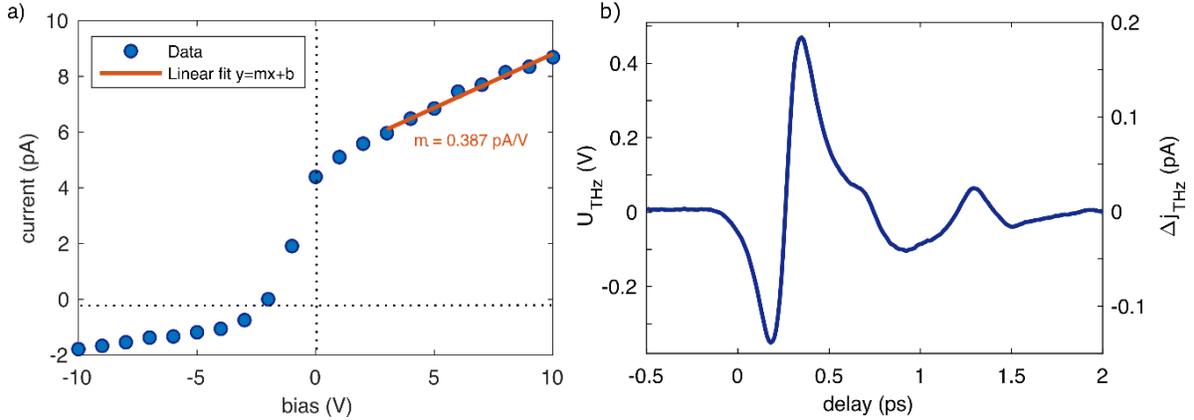

*Figure S.1* (a) Photocurrent-voltage curve used to calibrate the THz bias in Figure 2 in the main manuscript. (b) Calibrated THz bias (left y-axis) and corresponding THz-induced photocurrent change (right y-axis) measured at 8 V DC bias. ($d = 1$ µm and $I_{peak} = 0.6x10^{11}$ W/cm²)

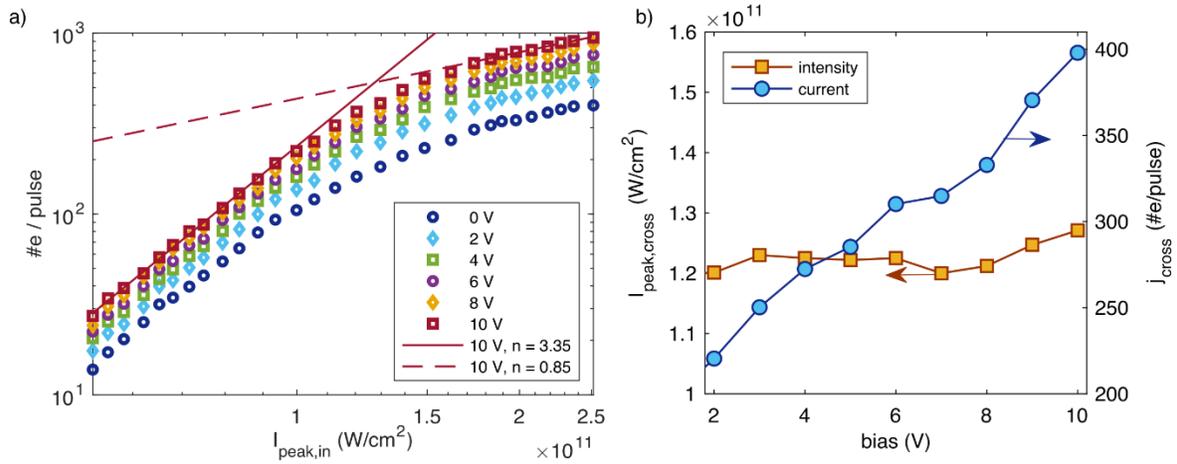

***Figure S.2*** *(a) Photocurrent versus NIR peak intensity at different DC bias measured at 1 μm tip-sample distance. Solid and dashed red lines are power law fits to the low- and high-power regimes shown exemplary for 10 V DC bias. (b) Laser intensity (left y-axis) and photocurrent (right y-axis) at which the transition from multiphoton scaling to nearly linear scaling occurs, defined as the crossing point between the power law fits of the two regimes (solid and dashed lines in (a)).*

## 2. Bias dependence of photocurrent-power scaling at large gap size

Figure S.2a) shows the dependence of the photocurrent on the incident laser intensity for DC bias in the range 0 V to 10 V measured at $d = 1$ μm for the tip condition used in Figure 2. We find that the transition from multiphoton scaling (power exponent $n \sim 3.4$) to a regime of nearly linear photocurrent scaling ($n \sim 0.9$) occurs at the same laser intensity for all DC bias, despite a 2x increase of the photocurrent from 0 V to 10 V, as plotted in Figure S.2b). This proves that space charge is not responsible for the observed saturation, but that we observe the transition to the strong-field regime[1]. This is reasonable considering the employed laser intensity and a Keldysh parameter of $\gamma \sim 2$ at the transition.

## 3. THz waveforms for a tip operating solely in the weak-field regime

It was argued recently that delayed photocurrents can originate from laser-driven inelastic re-scattering and re-emission of photoelectrons from a laser-excited tungsten tip[2]. In order to generate sufficiently delayed photocurrents, a significant amount of photoelectrons need to be driven back to the tip, re-enter the tip, undergo scattering processes inside the tip, and get re-emitted after a certain time delay. Especially in the presence of a static bias, the oscillating laser field needs to be strong enough to reverse an electron's trajectory and steer it back into the tip within less than an optical cycle. Hence, if operative, such source for delayed photocurrents can be expected to occur only in the strong-field regime of photoemission. We can exclude that electron scattering and subsequent re-emission is responsible for the observed THz waveform deformations, as for some tips we observe delayed photocurrents solely in the multiphoton photoemission regime. Figure S.3a) and S.3b) show the dependence of the photocurrent and the THz waveforms on the laser pulse energy for a blunter tip condition, respectively. In this case, strong THz waveform deformations of similar character as those in Figure 2 are observed purely in the multiphoton regime, as a nonlinear power scaling with constant slope $n = 2.24$ is observed in the investigated laser power range. This data is recorded from the same tip as used in Figures 2 and S.2, but after the tip condition has changed. We observe that (i) the photocurrent and its power scaling, (ii) the observation of a delayed photocurrent and the required threshold laser intensity, and (iii) the calibrated THz bias amplitude all correlate with the detailed nanoscale tip condition. This indicates that the nanoscale shape of the tip does not only affect the optical field enhancement but also the electron dynamics inside the apex considerably.

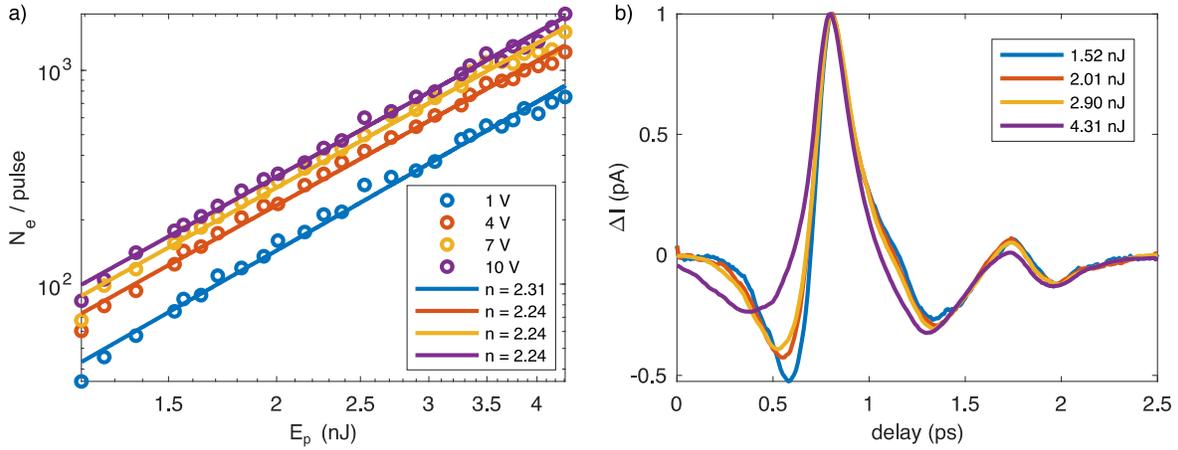

***Figure S.3*** *(a) Dependence of the photocurrent on the NIR pulse energy for a blunter tip, exhibiting multiphoton scaling for the entire pulse energy range. (b) Corresponding THz waveforms measured at four different powers (8 V DC bias, 1 μm tip-sample distance).*

## 4. Theoretical model

THz waveforms are calculated by the following procedure: First, we calculate the enhanced near-infrared (NIR) laser field and absorbed power density in the tip-sample junction using
5

the RF-Module of COMSOL Multiphysics 5.6 by solving the time-harmonic wave equation in three dimensions. The tip-sample junction is excited by an incident NIR Gaussian laser beam with 2 μm waist and peak laser field $F_{\text{in}}$, whose magnitude is varied in a range close to the incident NIR laser field strength used in the experiments.

Second, we convolve the obtained optical power density inside the tip, $S_a$, with the ballistic transport range $\lambda_b$ of electrons in tungsten.[3] Specifically, we convolve the absorbed power profile $S_a$ with an exponentially decaying impulse response function[4] $\propto e^{-\lambda_b r}/r$, which yields the power density distribution after spatial redistribution of energy due to ballistic transport, $S_b$. This is performed within the COMSOL Multiphysics simulation environment inside the 3D geometry of the tip by solving the Helmholtz equation

$$\nabla^2 S_b - \lambda_b^2 S_b = S_a. \qquad (S1)$$

Figures S.4.1a) and S.4.1b) compare the spatial profiles of $S_a$ and $S_b$ for tip-sample distances of $d = 10$ nm and $d = 1$ nm, respectively. The effect of ballistic transport on the spatial profile of the absorbed power density is clearly evident from the comparison of $S_a$ and $S_b$, which shows how ballistic transport spreads the energy deposited quasi-instantaneously inside a nanolocalized volumes such as an STM tip excited by an ultrafast laser pulse.

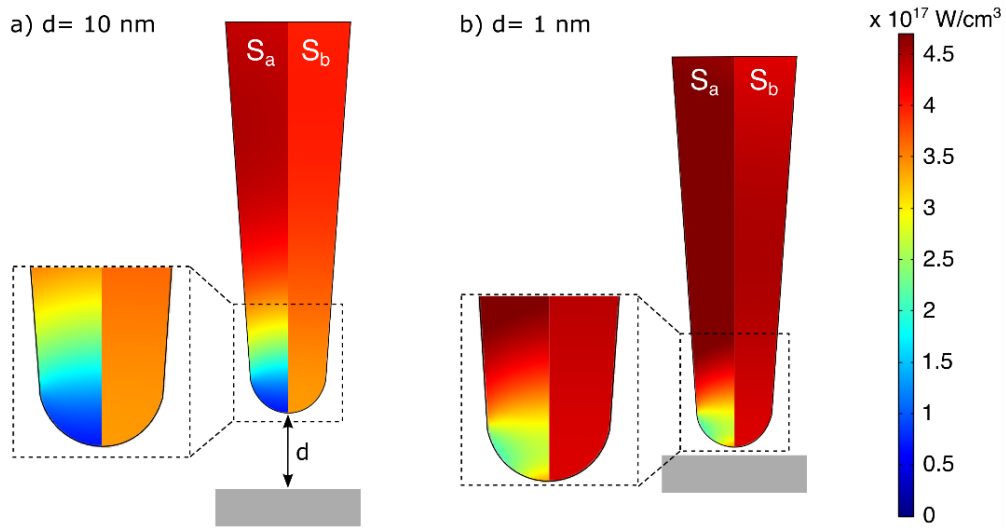

**Figure S.4.1** *Spatial profiles of the absorbed power density inside the STM tip before ($S_a$) and after ($S_b$) spatial re-distribution of energy due to ballistic electron transport for tip-sample distances of (a) 10 nm and (b) 1 nm.*

Third, we obtain the temperature evolution of the electronic and phononic sub-systems from the two-temperature model (TTM) by solving the differential equations

$$C_{el}\frac{\partial T_{el}}{\partial t} = \nabla(k_{el}\nabla T_{el}) - G(T_{el} - T_{ph}) + S_{TTM}(t) \qquad (S2a)$$

$$C_{ph}\frac{\partial T_{ph}}{\partial t} = G(T_{el} - T_{ph}) \qquad (S2b)$$

in three dimensions, where $T$ is temperature, $C$ is the specific heat capacity, $k$ is the thermal conductivity, $G$ is the electron-phonon coupling and the subscripts 'el' and 'ph' denote the electron and phonon sub-systems, respectively. $C_{el}$ and $k_{el}$ are considered temperature dependent and described by $C_{el}(T_{el}) = \gamma T_{el}$ and $k(T_{el}) = k_{eq}(T_{el}/T_{ph})$, where $\gamma$ is the electron heat capacity constant and $k_{eq}$ is the electron thermal conductivity at equilibrium. Since it can be assumed that ballistic redistribution of the energy takes place quasi-instantaneously on the time scale of the THz field, we use the power density profile $S_b$ to calculate the source term for the TTM,

$$S_{TTM}(t) = S_b \exp\left[-2\frac{t-t_0}{t_p}\right]^2, \qquad (S3)$$

where $t_p$ is the NIR laser pulse duration. The system of partial differential equations is solved numerically using the following parameter values:[4] $\gamma = 137.3 \text{ JK}^{-2}\text{m}^{-3}$, $k_{eq} = 150 \text{ Wm}^{-1}\text{K}^{-1}$, $G = 7.5 \times 10^{17} \text{ WK}^{-1}\text{m}^{-3}$, and $C_{ph} = 2.58 \times 10^6 \text{ JK}^{-1}\text{m}^{-3}$, which yields the electron and phonon temperatures at any point inside the tip.

Last, the calculated electronic temperature evolution at the tip apex is used to simulate the emission of electrons using Eqs. (2) – (5) from the main manuscript. The DC field at the apex, $F_{dc}$, is obtained from electrostatic simulations of the DC electric field inside the biased tip-sample junction using the AC/DC module of COMSOL Multiphysics 5.6, where the same tip geometry is used for DC and RF simulations. The peak THz field is determined from the experimentally known ratio of DC bias and THz bias and the simulated DC field as $F_{THz} = [U_{THz}/U_{dc}]F_{dc}$. Finally, we calculate the THz-induced change of the ultrafast thermionic current, $\Delta j_{th}(\tau)$, as a function of NIR-THz time delay using Eq. (2).

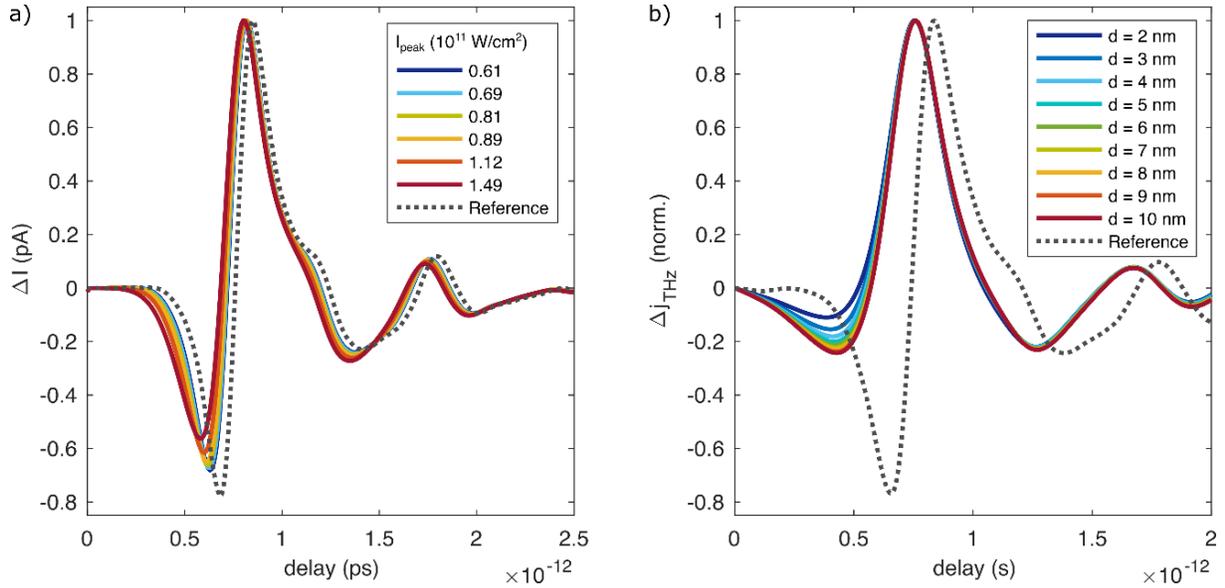

***Figure S.4.2**  THz waveforms calculated for pure thermionic emission and their dependence on (a) incident laser intensity ($d = 1\ \mu m$) and (b) tip-sample distance ($E_{in} = 1.4$ V/nm and $I_{peak} = 2.6 \times 10^{11}$ W/cm², respectively) at 8 V DC bias.*

Figure S.4.2a) shows the dependence of the THz waveform, which is simulated for pure thermionic currents, on the NIR laser intensity without contributions from nonthermal current due photoemission or photoassisted tunneling. The simulations reveal that 'pure thermionic' THz waveforms do not reproduce the original THz waveform exactly. Even at comparably low electron temperatures, at which the thermionic current density is very small, the calculated THz waveform exhibits typical low-pass filtered character, where the deformations of its shape originate predominantly from the slow decay, and the temporal shift from the non-instantaneous increase of the electron temperature and the corresponding transient thermionic current.

Figure S.4.2b) shows THz waveforms simulated for pure thermionic currents at varying gap size. As expected from the distance dependence of the enhanced laser intensity and thus of the ultrafast heating of the electrons, which both increases inversely with gap size, the THz waveforms become more distorted at smaller gap size. This trend is further enhanced by the reduced barrier width at small gap sizes.

## 5. SEM image of STM tip

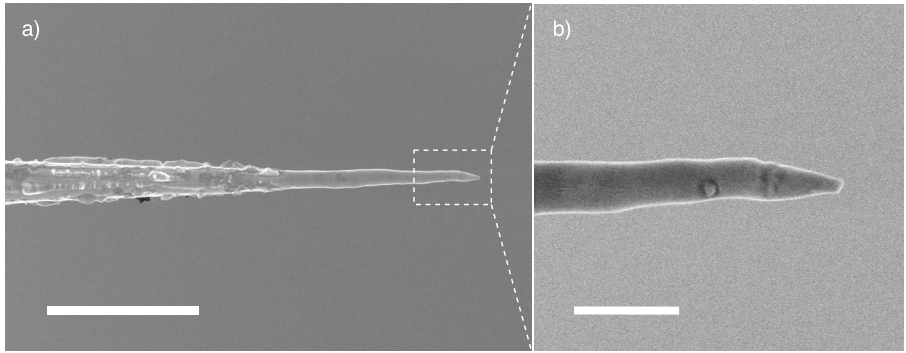

*Figure S.5 SEM images of the tungsten tip used in this work. Images are recorded after several weeks of tip usage inside the STM. Scale bars correspond to (a) 3 μm and (b) 500 nm. The area of the tip shaft with reduced surface roughness in (a) matches the approximate laser spot size.*

## 6. Power-dependent current-distance curves

In order to extract the nonlinearity of the photocurrent at nanometer gap distances, we measure photocurrent-distance curves at different laser powers. Figure S.6.1a) shows few examples of $j_{ph} - z$ curves over the power range used here. For all powers the curves overlap at the closest distances $\Delta z \lesssim 1.5$ nm and we observe a clear transition to the DC tunneling regime, even at the highest powers. At the set point, the current of 1 nA is an order of magnitude larger than the photocurrent observed outside but close to the DC tunneling range. We can thus assume that the set point distance is determined predominantly by the DC tunneling current. Fitting the DC current and subtracting this from the total current thus yields the distance scaling of the pure photocurrent. Figure S.6.1b) shows the laser power dependence of the resulting photocurrent and linear fits for distances in the range 1 nm to 12 nm. The retrieved nonlinearities are plotted versus relative gap distance in Figure 4c) in the main manuscript.

Even though at the set point the photocurrent contribution to the total current is small, the absolute tip position and gap distance will depend on the NIR laser power, as the feedback will compensate for the small but increasing photocurrent contribution at higher laser intensities. To estimate this effect, we extrapolate the fitted DC current at each power to the quantum conductance $G_0$, which reveals a difference of the absolute gap distance of ~0.5 nm between the highest and lowest power. Figure S.6.2a) shows the power dependence of the estimated absolute tip-sample distance $d$, revealing the set point correction due to the photocurrent contribution to the total current. Note that this is only a small fraction of the measured change in $z$ versus power, plotted in Figure S.6.2b), which originates predominantly from thermal expansion. We checked that the 0.5 nm offset due to the photocurrent has negligible effect on the extracted nonlinearity in the investigated range at our conditions. We estimate this by shifting the $j_{ph} - z$ curves along the $\Delta z$-axis according to their difference in the absolute distance $d$. Figure S.6.2c) compares the effective nonlinearity retrieved from the DC-fit corrected, but unshifted $j_{ph} - z$ curves (blue, same as Figure 4c)) and the DC-fit corrected, but shifted and $d$-corrected curves (orange). We find that the ~0.5 nm change of the absolute tip-sample distance has insignificant effect on $n_{\text{eff}}$ in the range $\Delta z > 1$ nm at our conditions. We are hesitant to conclude on the range $\Delta z < 1$ nm as this is difficult to fit reliably due to the large scatter of the data in the current-power curves originating from subtraction of the fitted DC current. It is clear, though, that the effective nonlinearity decreases significantly and approaches the range $n \sim 1$ at the closest distances. This corroborates the assumption that a significant amount of electrons in channel $n = 1$ at low energies contributes to the photocurrent and is responsible for the very fast transition to the original THz waveform within a few Angstrom distance change.

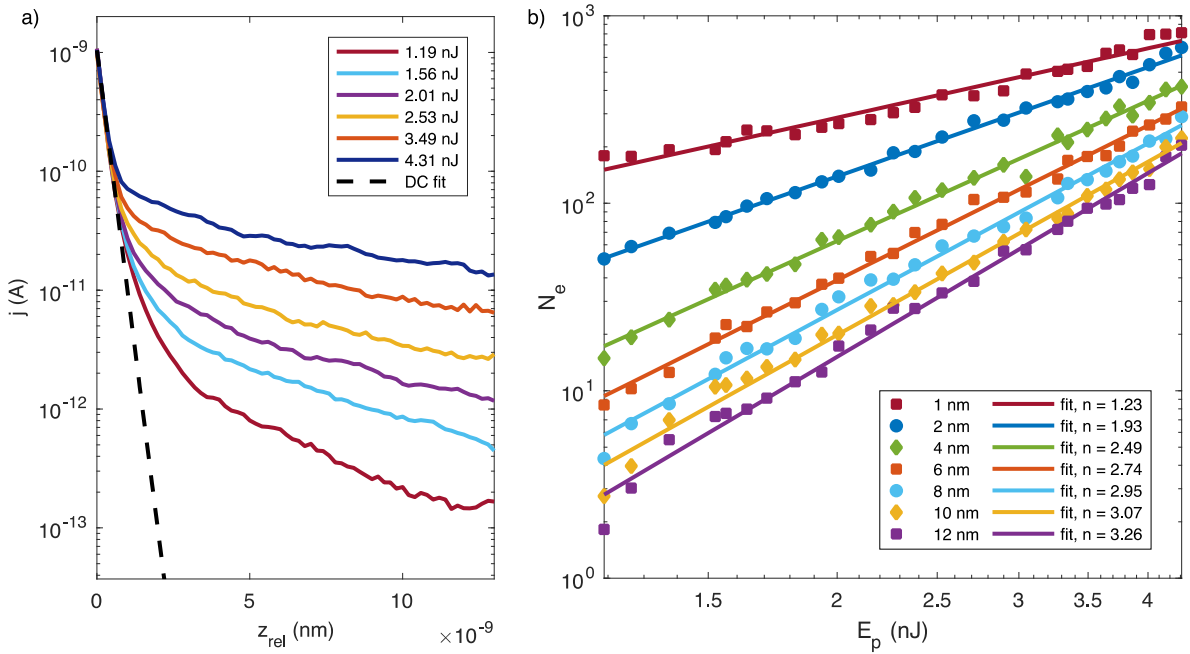

*Figure S.6.1 (a) Photocurrent-distance curves measured at different NIR laser powers. Zero relative distance is defined by the STM set point of 1 nA and 10 V. The black dashed line is a fit to the DC tunneling current, which dominates at small distances. (b) Laser power scaling of the photocurrent at different gap distances, as retrieved from vertical cuts of the $I_{ph} - z$ curves in (a) after subtraction of the DC current from the total current. Solid lines are power law fits to extract the nonlinearity $n$.*

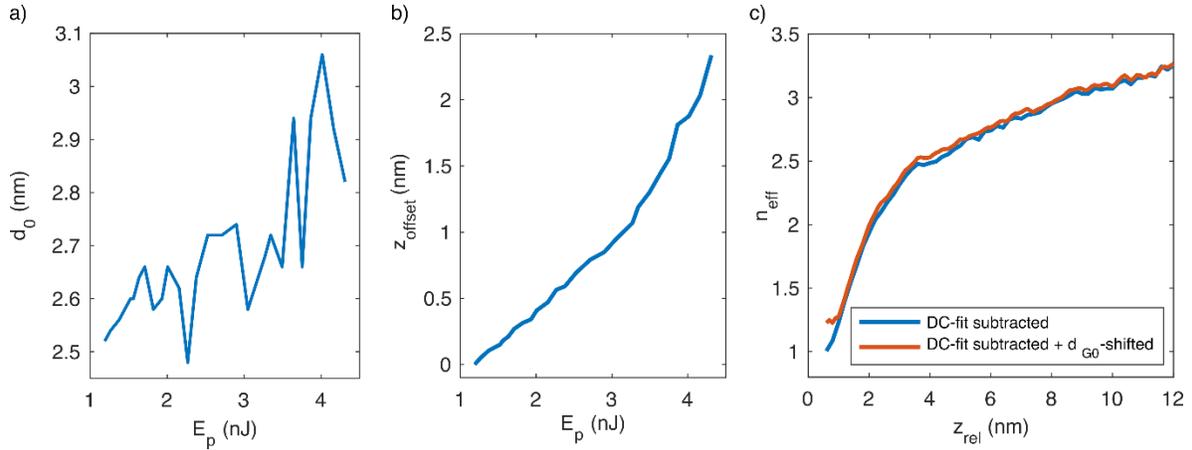

***Figure S.6.2*** *(a) Estimated absolute gap distance at the set point as a function of laser pulse energy, obtained by extrapolating the fitted DC current to the quantum conductance $G_0$. The contribution of the photocurrent to the total current results in ~0.5 nm variation of the gap distance at the set point between the lowest and highest powers used. (b) Offset of the z-piezo versus incident pulse energy, which originates predominantly from steady-state thermal expansion of the tip, which is significantly larger than the z-correction due to an increasing contribution of photocurrent to the total current. (c) Change of effective nonlinearity with gap size. Blue curve: only the DC current is subtracted from the $I_{ph} - z$ curves (Figure S.6.1a); Red curve: $I_{ph} - z$ curves are additionally shifted by the gap distance change shown in (a) to account for the z-correction at the set point due to the increase of the photocurrent with increasing laser power.*